\documentclass[twocolumn,showpacs,superscriptaddress,prl]{revtex4-1}

\usepackage[dvips]{graphicx}	
\usepackage{dcolumn, bm, color, amsmath}
\usepackage{amssymb}
\usepackage{tikz}
\usepackage{colortbl} 
\usepackage{array}          

\newcommand{\ffullhexagonSymbolOne}{
  \tikz[baseline=0.07cm]{
    \draw (0,0) -- (0.2,0) -- (0.3,0.15) -- (0.2,0.3) -- (0,0.3) -- (-0.1,0.15) -- cycle;
    \fill (0.2,0.3) circle (1.5pt);  
    \fill (0, 0.3) circle (1.5pt);    
    \fill (0.2,0) circle (1.5pt);    
    \fill (0,-0.00375) circle (1.5pt);
    \fill (0.3,0.15) circle (1.5pt);  
    \fill (-0.1,0.15) circle (1.5pt); 
  }
}

\newcommand{\ffullhexagonSymbolTwo}{
  \tikz[baseline=0.07cm]{
    \draw (0.1,0.3) -- (0.25,0.225) -- (0.25,0.075) -- (0.1,0.001875) -- (-0.05,0.073125) -- (-0.05,0.225) -- cycle;
    \fill (0.1,0.3) circle (1.5pt);    
    \fill (0.1,0.001875) circle (1.5pt);   
    \fill (0.25,0.225) circle (1.5pt);     
    \fill (-0.05, 0.225) circle (1.5pt);      
    \fill (0.25,0.075) circle (1.5pt);       
    \fill (-0.05,0.073125) circle (1.5pt);         
  }
}
\newcommand{\hexagonSymbolOne}{%
  \tikz[baseline=0.07cm]{
    \draw (0,0) -- (0.15,0) -- (0.225,0.15) -- (0.15,0.3) -- (0,0.3) -- (-0.075,0.15) -- cycle;
    \fill (0.15,0.3) circle (1.5pt);  
    \fill (0, 0.3) circle (1.5pt);      
    \fill (0.15,0) circle (1.5pt);      
    \fill (0,-0.00375) circle (1.5pt);  
  }
}
\newcommand{\hexagonSymbolTwo}{%
  \tikz[baseline=-0.004cm]{
    \draw[rotate=90] (0,0) -- (0.15,0) -- (0.225,0.15) -- (0.15,0.3) -- (0,0.3) -- (-0.075,0.15) -- cycle;
    \fill[rotate=90] (0.225,0.15) circle (1.5pt); 
    \fill[rotate=90] (-0.075,0.15) circle (1.5pt); 
  }
}

\begin{document}

\title{Investigating the superconducting state of 2$H$-NbS$_2$ as seen by the vortex lattice}

\author{A. Alshemi}
\email{ahmed.alshemi@sljus.lu.se}
\affiliation{Division of Synchrotron Radiation Research, Department of Physics, Lund University, SE-22100 Lund, Sweden}
\author{E. Campillo}
\affiliation{Division of Synchrotron Radiation Research, Department of Physics, Lund University, SE-22100 Lund, Sweden}
\author{E.~M.~Forgan}
\affiliation{School of Physics and Astronomy, University of Birmingham, Edgbaston, Birmingham, B15 2TT, UK}
\author{R. Cubitt}
\affiliation{Institut Laue Langevin, 71 Avenue des Martyrs, F-38000 Grenoble cedex 9, France}
\author{M. Abdel-Hafiez}
\affiliation{Department of Applied Physics and Astronomy, University of Sharjah, P. O. Box 27272 Sharjah, United Arab Emirates} 
\affiliation{Department of Physics and Astronomy, Uppsala University, Box 516, SE-75120 Uppsala, Sweden}
\author{E.~Blackburn}
\affiliation{Division of Synchrotron Radiation Research, Department of Physics, Lund University, SE-22100 Lund, Sweden}

\begin{abstract}
 
2$H$-NbS$_2$ is a classic example of an anisotropic multi-band superconductor, with significant recent work focussing on the interesting responses seen when high magnetic fields are applied precisely parallel to the hexagonal niobium planes.  It is often contrasted with its sister compound 2$H$-NbSe$_2$ because they have similar onset temperatures for superconductivity, but 2$H$-NbS$_2$ has no charge density wave whereas in 2$H$-NbSe$_2$ the charge density wave order couples strongly to the superconductivity.  Using small-angle neutron scattering, a bulk-sensitive probe, we have studied the vortex lattice and how it responds to the underlying superconducting anisotropy.  This is done by controlling the orientation of the field with respect to the Nb planes.  The superconducting anisotropy, $\Gamma_{ac} = 7.07 \pm 0.2$, is found to be field independent over the range measured (0.15 to 1.25 T), and the magnetic field distribution as a function of the applied magnetic field is found to be in excellent quantitative agreement with anisotropic London theory modified with a core-size cut-off correction, providing the first complete validation of this model.  We find values of $\lambda_{ab} = 141.9 \pm 1.5 $ nm for the in-plane London penetration depth, and $\lambda_{c} \sim$ 1 \textmu m for the out-of-plane response.  The field-independence indicates that we are primarily sampling the larger of the two gaps generating the superconductivity in this material.

\end{abstract}


\date{\today}

\maketitle

\section*{Introduction}
 
Recent studies of the superconducting phase diagram of 2$H$-NbS$_2$ \cite{Cho2021} and 2$H$-NbSe$_2$ \cite{Wan2023} have revealed that the superconducting state in these sister materials is very sensitive to the orientation of the magnetic field with respect to the basal plane.  Different types of spatially textured superconductivity \cite{FF,LO} have been conjectured at high magnetic fields in both materials. 

These two compounds belong to the larger family of transition metal dichalcogenides (TMDCs).  This is a class of highly 2D materials, with hexagonal planes of transition metals weakly coupled along the \textbf{c} axis, with members displaying multiple types of electronic order, including charge density wave (CDW) order, Mott-insulating behaviour and superconductivity \cite{Manzeli2017}. 2$H$-NbSe$_2$ is a classic example of the interaction between a charge density wave state and superconductivity, and the family 2$H$-MX$_{2}$ (M = Nb, Ti, Ta; X = S, Se) all have similar electronic band structures in the normal state. When they become superconducting, typically a larger gap develops on Fermi surface sheets with a more 2D nature, and the smaller gap appears on Fermi surface sheets with a more 3D character. Within this family, 2$H$-NbS$_2$ stands out because it is the only one in which CDW order has not been seen in the bulk (see Table I); in the literature there is some dispute if it exists in the monolayer 1$H$-NbS$_2$ \cite{Knispel2024-monolayerNbS2,Lin2020}. This means that comparing 2$H$-NbS$_2$ and 2$H$-NbSe$_2$ is a clean way to check out how the CDW affects the physics observed in these materials. 

\begin{figure*}
\begin{center}
    \includegraphics[width=1\linewidth]{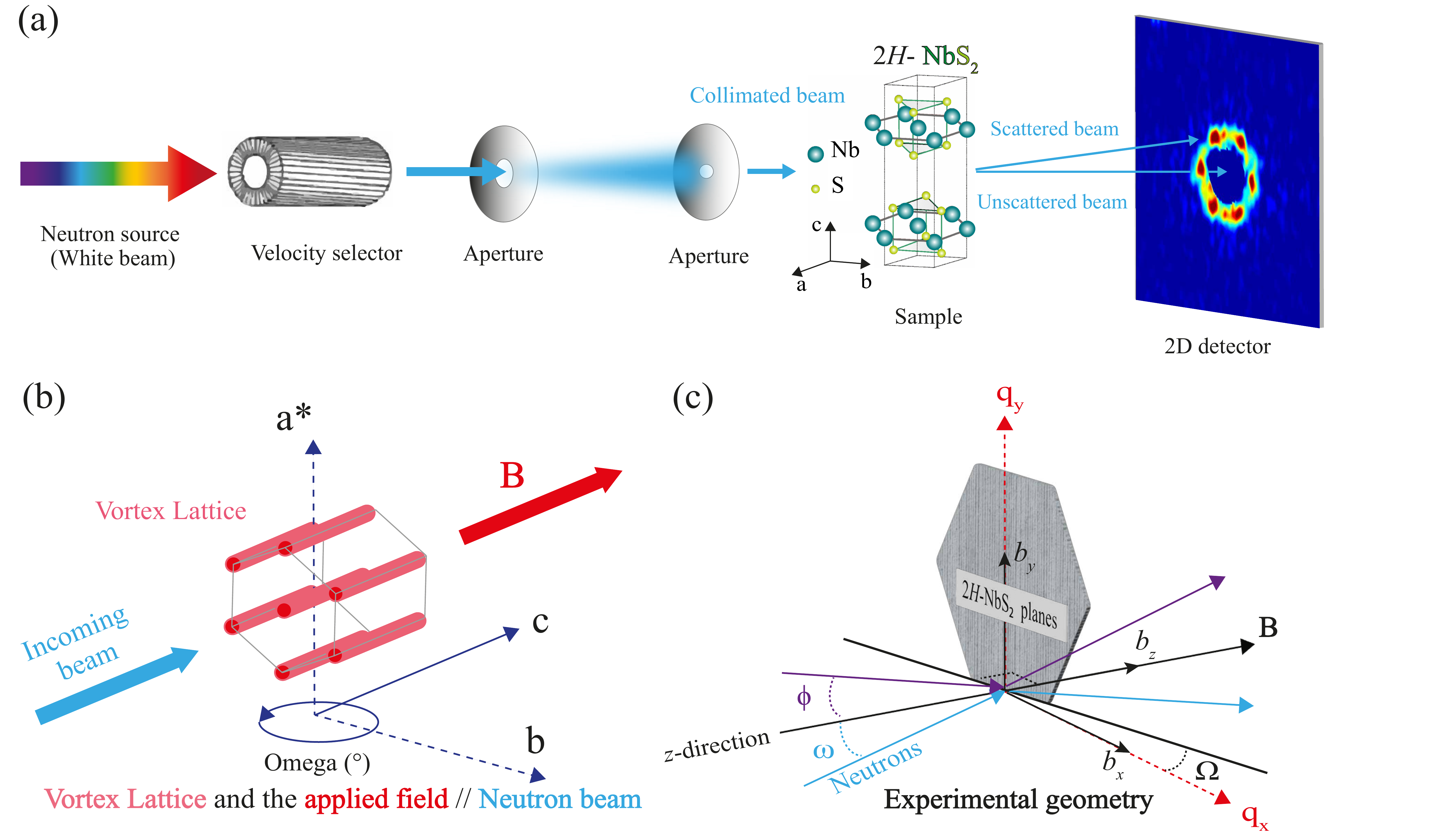}
\end{center}
	\caption{\textbf{Schematic diagram of the experimental setup.} (a) Typical small-angle neutron scattering (SANS) instrument setup for diffraction by a superconducting vortex lattice.  The neutrons pass through a velocity selector, which sets their average wavelength (usually with a full-width half-maximum (FWHM) spread $\sim 10\% $). The beam is well collimated, with a long evacuated flight path before and after the sample to minimize air scattering.  The transmitted main beam is caught on a neutron-absorbing beamstop to allow the Bragg reflections, which are scattered at small angles, to be visible on a 2D multidetector. (b) Sketch of the `parallel' field geometry used in the experiment; here the individual vortices are aligned parallel to the applied field, and then by rotating the magnet and sample together through angles $\Omega$ or $\phi$, the Bragg condition for the vortex lattice can be met. When the data obtained at different angles are summed together, an image such as that shown on the multi-detector will be seen. (c) Experimental geometry. The coordinate system is defined with the $z$ direction along $\textbf{B}$ and the components of observed VL scattering vectors are denoted as $q_{x}$ and $q_{y}$. The applied magnetic field B is rotated away from the 2$H$-NbS$_2$ plane by an angle $\Omega$ and the longitudinal and transverse components of the field modulation are denoted by $b_{z}$ and $b_{x} ~ \& ~ b_{y}$, respectively.}
		\label{Graph1}
\end{figure*}

\newcolumntype{M}[1]{>{\centering\arraybackslash}p{#1}}
\begin{table}[h]
\begin{center}
\renewcommand{\arraystretch}{1.5} 
    \begin{tabular}{M{0.15\textwidth} | M{0.15\textwidth} | M{0.15\textwidth}}
     \hline\hline
     \rowcolor{gray}
      \textcolor{white}{\textbf{TMDC}} & \textcolor{white}{$T_{\mathrm{CDW}}$ (K)} & \textcolor{white}{$T_c$ (K)} \\
      \hline
      2$H$-NbSe$_2$ & 33.5 & 7.3  \\ \hline
      2$H$-NbS$_2$ & none & 5.5  \\ \hline
      2$H$-TaSe$_2$ & 122.3 & 0.15 \\ \hline
       2$H$-TaS$_2$ & 78 & 0.8  \\ 
      \hline
      \hline
    \end{tabular}
    \newline
\caption{\textbf{Transition temperatures for the charge density wave and superconducting states in selected transition metal dichalcogenides.}  Values are taken from Refs.~\cite{Moncton1977,Wilson1975,AbdelHafiez2016}.}
\end{center}
\end{table}

Both materials are considered to be excellent examples of two-band superconductors with two different $s$-wave  gaps, from scanning tunnelling spectroscopy \cite{Guillamon2008}, specific heat \cite{Kacmarcik2010,Kobayashi_1977} and Andreev reflection \cite{Majumdar_2020} studies.  For both materials, the Fermi surfaces are similar, with three types of Fermi surface sheet.  The Fermi surface sheets that arise from the Nb 4$d$ bands and exhibit superconductivity are cylinders centred around the $\Gamma$ and K points in the Brillouin zone. They have different levels of corrugation leading to more 3D character in those around the $\Gamma$ point.  There is also a (non-superconducting) smaller pancake-like sheet at $\Gamma$ associated with the chalcogen.  This has been reported by many independent groups; a nice description is given by Noat \emph{et al.}~\cite{Noat_2015}.  Where there are two superconducting sheets, one may expect different gap magnitudes and superconducting anisotropies, although the effective vortex core radii (related to the coherence lengths) are expected to lock together in most cases \cite{Ichioka_2017}. 

In this context, we use the vortex lattice (VL) that develops in the superconducting state to probe the superconducting response, with a particular eye on the effects of anisotropy.  In 2$H$-NbSe$_2$, the vortex lattice has been studied using a variety of methods.  Two results stand out.  Firstly, the individual vortex cores have a six-fold star-shaped structure \cite{Guillamon2008}, reflecting the symmetry of the CDW order. Secondly, the vortex lattice consists of one hexagonal domain; to a first approximation the vortices lie parallel to the external field (see Fig.~\ref{Graph1}b). On rotating the magnetic field towards the basal plane, this domain distorts, reflecting the underlying effective mass anisotropy. However, the unit cell vectors of this domain do not change direction and are, in fact, pinned to the crystallographic $\textbf{a}^*$ axis \cite{Hess1994,Gammel1994}.  This unexpected observation indicates that in 2$H$-NbSe$_2$ the orientation of the VL is not in agreement with the predictions of anisotropic London theory, as seen in, for example, YBa$_2$Cu$_3$O$_{7-\delta}$ \cite{Kealey_2001}, Sr$_2$RuO$_4$ \cite{Rastovski2013} and KFe$_2$As$_2$ \cite{KuhnKFA}.

In contrast, there are relatively few direct studies of the vortex lattice in 2$H$-NbS$_2$, with a scanning tunneling microscopy and spectroscopy study by Guillamon \textit{et al.}~\cite{Guillamon2008} confirming that a well-ordered vortex lattice can be seen at the sample surface, and that the vortex core has a standard circular shape. 

\begin{figure*}[ht]
\begin{center}
	\includegraphics[width=1\linewidth]{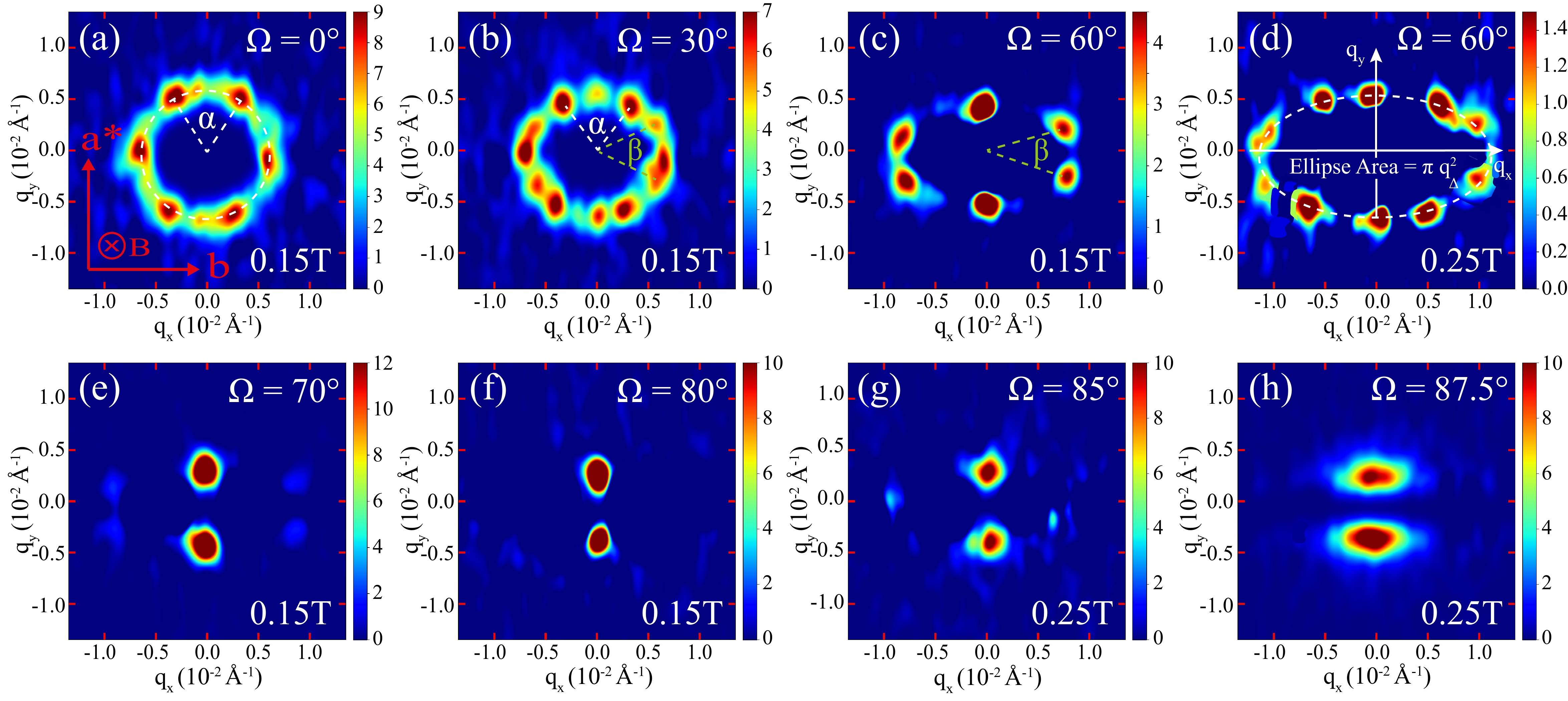}
\end{center}
	\caption{\textbf{SANS diffraction pattern of the VL in 2$H$-NbS$_{2}$ as a function of field rotation angle $\boldsymbol{\Omega}$}.  We show the results of measurements made at 1.5 K for seven different field angles, $\Omega$, at either 0.15 T or 0.25 T, depending on the closeness of the Bragg spots to the direct beam. The images are sums over $\phi$ rocking scans about the horizontal axis perpendicular to the incoming beam, minus backgrounds. The direct beam has been masked off in software. The white and green dashed lines in panels (a), (b) and (c) represent the opening angles of $\alpha$ and $\beta$ for Domain I (\protect\ffullhexagonSymbolOne) and Domain II (\protect\ffullhexagonSymbolTwo), respectively. The white dashed ellipse depicted in panel (d) lies on top of the Bragg reflections from the distorted VL. It has a major-to-minor axis ratio $\Gamma_{\mathrm{VL}}$ and the reciprocal-space area of the ellipse is given by $\pi q_{\vartriangle}^2$, with $q_{\vartriangle}=2 \pi\left(2 B / \sqrt{3} \Phi_0\right)^{1 / 2}$.} 
	\label{Graph2}
\end{figure*}

Here we present a study of the vortex lattice as measured deep inside the superconducting state at 1.5 K, extracting information on the superconducting anisotropy, the penetration depth $\lambda$ and the coherence length $\xi$. This is done by neutron diffraction from the magnetic field distribution associated with the vortex lattice, whereby the Fourier components of this periodic distribution can be extracted from the resulting Bragg reflections.  A schematic of the experimental setup is shown in Fig.~\ref{Graph1}.  This is done in the small-angle scattering regime because the inter-vortex distances are much larger than available neutron wavelengths.  To probe the anisotropy of the system, we create vortex lattices with different angles $\Omega$ between the applied magnetic field and the $c$ axis. Full details of the experimental protocols are given in the Methods.  Fig.~\ref{Graph2} shows a range of diffraction patterns collected at $T$ = 1.5 K and magnetic fields ranging from 0.15 T to 0.25 T, for a range of values of $\Omega$, running from 0$^{\circ}$ to 87.5$^{\circ}$, prepared using the field cooling process described in the Methods.  It is clear that rotating the field from being parallel to $\textbf{c}$ axis ($\Omega = 0^{\circ}$) to nearly being in the basal plane ($\Omega = 87.5^{\circ}$) distorts the hexagonal vortex lattice, indicating the anisotropic nature of 2$H$-NbS$_{2}$.

\noindent
\section*{Results and Discussion}

\subsection*{Vortex lattice structure}

When the field ($B$ = 0.15 T) is parallel to $\textbf{c}$ axis ($\Omega = 0^{\circ}$), the diffraction pattern is perfectly hexagonal (Fig.~\ref{Graph2}a), with the Bragg peaks appearing at $q_{\vartriangle}=5.6(1) \times 10^{-3}$ \AA$^{-1}$, as compared with the ideal value for a hexagonal VL $q_{\vartriangle}=2 \pi\left(2 B / \sqrt{3} \Phi_0\right)^{1 / 2}=5.7 \times 10^{-3}$\AA$^{-1}$ at $B$ = 0.15 T.  $\Phi_0  = h/2e$ is the flux quantum. Because of the six-fold crystal symmetry, one of two hexagonal domains separated by 30$^{\circ}$ might be expected to be energetically favoured. We observe that one domain (Hexagonal Domain I [\ffullhexagonSymbolOne]) is dominant, with trace amounts of Hexagonal Domain II [\ffullhexagonSymbolTwo]. Prior to changing $\Omega$, the diffraction spots of Domain I lie parallel to the $\textbf{a}/\textbf{b}$ axes, indicating that, in real space, the vortex lattice planes are perpendicular to the Nb nearest neighbour directions.  When $\Omega = 0^{\circ}$, the $q_x$ and $q_y$ directions on the detector are probing directions at right angles within the basal plane, and we expect the superconducting parameters ($\lambda$ and $\xi$) to be essentially isotropic around this plane. 

At $\Omega = 30^{\circ}$ at 0.15 T, the ratio of the two domains has shifted, making Domain II easier to see, and we can also see the VL begin to distort due to the difference in the superconducting properties in- and out-of-plane.  The Bragg peaks move further away from the beam centre along the horizontal axis and closer along the vertical axis.  The same distortion applies to both domains, as can be seen by considering that all of the Bragg spots lie on the same ellipse (Fig.~\ref{Graph2}b,d).  

\begin{figure}[t]
        \includegraphics[width=\columnwidth]{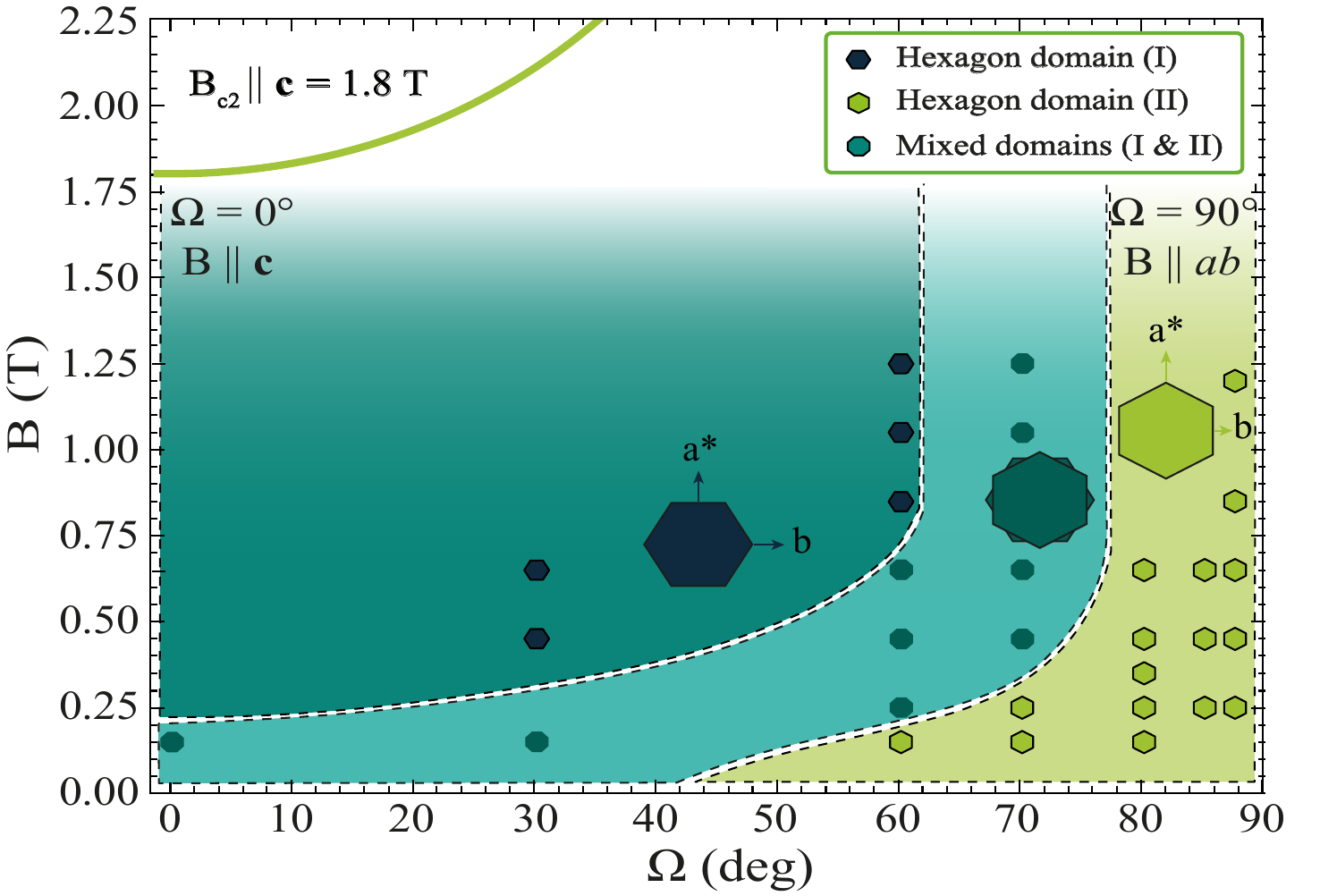}
	\caption{\textbf{The B-$\Omega$  phase diagram of the VL in 2$H$-NbS$_{2}$ at $T$ = 1.5~K.} The shaded areas are represented by a gradient of green color, transitioning from dark green to pale green, signifying the change in VL structure as the direction of the magnetic field {\textbf B} is changed relative to the crystallographic $ab$-plane. Hexagon symbols represent distinct vortex lattice domains, with dark blue hexagons for domain I (\protect\ffullhexagonSymbolOne), light green for domain II (\protect\ffullhexagonSymbolTwo), and a teal dodecagon indicating regions with mixed domains (I $\&$ II).  The dashed black lines define the transitions between these domains as a function of $B$ and $\Omega$. The upper critical field ($B_{\text{c2}}$) line is the parameterisation from the ratio of $B_{c 2}^{\text {orb }}(\theta) / B_{c 2 \| a b}^{\text {orb }}$ using values from Ref.~\cite{Cho_2022}.  $B_{c 1  \| c}$ is estimated to be $\sim$ 30 mT \cite{Majumdar_2020}.}
	\label{Graph3}
\end{figure}

As the field is rotated further, Domain II becomes the dominant form, and indeed is the only domain seen for $\Omega > 70^{\circ}$.  As the distortion increases, the top and bottom reflections become more prominent.  The side reflections also move out of the window captured by the 2D detector.  The regions in which the different domains are seen as a function of magnetic field and $\Omega$ are shown in Fig.~\ref{Graph3}. No intermediate structures are seen at any point. These observations suggest that the transition between Domains I and II is first order in character, as reported in CaAlSi \cite{CaAlSi-Biswas}.  

As $\Omega$ is increased, $q_y$ remains in the same orientation in the basal plane of the crystal lattice, but $q_x$  starts to mix basal plane and out-of-plane components. A VL structure with Bragg peaks pinned to the $q_y$ direction should then be preferred,  as observed in, for example, YBa$_2$Cu$_3$O$_{7-\delta}$ \cite{Kealey_2001}, Sr$_2$RuO$_4$ \cite{Rastovski2013} and KFe$_2$As$_2$ \cite{KuhnKFA}.  However, as noted above, this was \emph{not} observed in 2$H$-NbS$_2$'s sister compound, 2$H$-NbSe$_2$, presumably due to a stabilising effect associated with the CDW formation \cite{Gammel1994}.  

\begin{figure*}[t]
\includegraphics[width=1\linewidth]{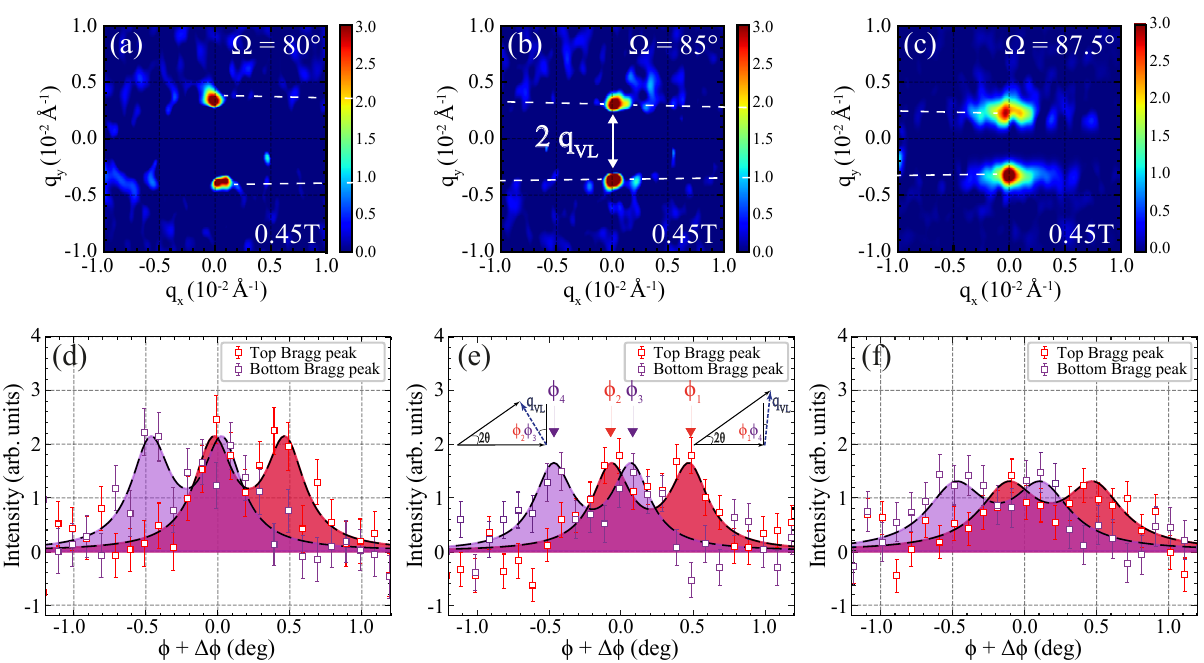}
	\caption{\textbf{Spin-splitting of the VL Bragg reflections due to spin-flip scattering.} Vortex lattice diffraction patterns as a function of rotation angle ($\Omega$) Vortex lattice rocking curves at 0.45 T and 1.5 K. Each rocking curve is fitted by two Lorentzian.  The VL anisotropy increases as $\Omega$ comes close to the basal plane, as indicated by the dashed white lines in the upper panels (a)-(c). Lower panels (d)-(f) represent the rocking curves corresponding to the diffraction patterns in the upper panels. The rocking curves show the scattered intensity distribution plotted as a function of the tilt angle deviation  ($\phi + \Delta \phi)$  relative to the rocking curve center which has slight zero offset at $\phi=0.09 ^{\circ}$. Two distinct peaks, indicative of Zeeman splitting from transverse field modulation (spin flip), the top Bragg reflection is represented in red, while the bottom Bragg reflection is depicted in purple. Both sets of peaks are fitted with two Lorentzians, as shown by the dashed lines, delineating the peak intensities and width for the top and bottom spots as a function of rotation angle ($\Omega$). The scattering geometry triangles are depicted in (e) (with angles exaggerated for clarity) for the two different SF processes: Spin-up to spin-down (L.H.S) and spin-down to spin-up (R.H.S), The scattering angle ( \(2\theta = 2\phi_0\)) is the same in both cases, but different tilt angles are required to satisfy the Bragg condition \(\phi_{2/3}\) (kinetic energy loss) and \(\phi_{1/4}\) (kinetic energy gain) for the top and bottom Bragg spots, respectively.}
		\label{Graph4}
\end{figure*}
\subsection*{Superconducting anisotropy}
\noindent
The vortex lattice becomes increasingly anisotropic as the field is rotated toward the basal plane, because the London penetration depth is different in the basal plane and along the $\textbf{c}$ axis.  This leads to the distortion of the diffraction pattern because the screening currents no longer follow circular paths close to the vortex core.  2$H$-NbS$_2$ falls into the class of uniaxial anisotropic superconductors,  which have a distortion of the hexagonal VL as a function of $\Omega$. The anisotropy of the vortex lattice, $\Gamma_{\mathrm{VL}}$ is defined as the ratio of the semi-major and semi-minor axes of the ellipse circumscribing the diffraction spots, as given by \cite{L.J.Cambell_PRB_1988}
\begin{equation}
\Gamma_{\mathrm{VL}} = \frac{\Gamma_{ac}}{\sqrt{\sin^2 \Omega+(\Gamma_{ac} \; \cos \Omega)^2}},
\label{gamma}
\end{equation}
where $\Gamma_{ac}$ corresponds to the $a-c$ anisotropy of the penetration depth. To evaluate $\Gamma_{ac}$, we therefore need to measure $\Gamma_{\mathrm{VL}}$ as a function of $\Omega$.  There are multiple ways to extract this information, and different ways have to be used at different angles, primarily because not all of the spots are measured at higher angles (Fig.~\ref{Graph2}). 

For $\Omega \leq 70^{\circ}$, the following methods were used:

\begin{itemize}
    \item The area of the Brillouin zone associated with the VL is fixed for a given value of field, and can be calculated directly for the perfect hexagon.  Given this, the position in reciprocal space of the spots, or, equivalently, the opening angles $\alpha$ and $\beta$ (marked in Fig.~\ref{Graph2}a-\ref{Graph2}c) can be used to calculate $\Gamma_{\mathrm{VL}}$, although the two domains require slightly different treatment. For Domain I (\ffullhexagonSymbolOne), $\Gamma_{\mathrm{VL}} = \sqrt{3} ~ [\tan (\alpha/2 )]$, or by taking the $q_y$ value for the spots    \hexagonSymbolOne, $\Gamma_{\mathrm{VL}} = 3/4[q_{\vartriangle}/q_{y}]^2$. For Domain II (\ffullhexagonSymbolTwo), we have  $\Gamma_{\mathrm{VL}} = [q_{\vartriangle}/q_{y}]^{2}$ where $q_y$ is measured for the spots \hexagonSymbolTwo , or $\Gamma_{\mathrm{VL}} = 1/(\sqrt{3}~[\tan (\beta / 2)])$.  
    
    \item Fitting an ellipse that meets the area constraint ($A_{ellipse} = \pi q_{\vartriangle}^2 = 8\pi^3  B/ \sqrt{3 \Phi_0}$, shown in Fig.~\ref{Graph2}d) to the six Bragg spot positions gives the semi-major and semi-minor axes, and hence $\Gamma_{\mathrm{VL}}$ directly. This method cannot be used for $\Omega > 70 ^{\circ}$ as only the top and bottom spots are visible.
\end{itemize}

For $\Omega > 70^{\circ}$, only the top and bottom Bragg spots of Domain II are visible, so the first of these methods is employed; in this case the position of the spots is $q_{\mathrm{VL}} = q_y$.  However, in anisotropic superconductors when the vortices are tilted away from a principal axis, the field distribution associated with the vortices develop transverse field components that vary with $\Omega$ \cite{S.Thiemann_1989}. (We later give expressions for these in Eqs.~\ref{eq:bx}, \ref{eq:by}, \ref{eq:bz}) Even if the near-horizontal spots at large angles did fall on the detector, they would have almost no intensity. This because the transverse components (labelled $b_x$ \& $b_y$ in Fig.~\ref{Graph1}c) are close to zero at small $q_y$ and the longitudinal component $b_z$ also falls off at large $\Omega$.  For the vertical spots, the transverse field component $b_x$ dominates and flips the spin of the scattering neutrons. Neutrons after spin flipping parallel or anti-parallel to the applied magnetic field will have slightly different Zeeman energies, leading to a change in the kinetic energy and hence velocity of the scattered neutron. This results in the Bragg spots splitting into two separate peaks in the $\phi$ rocking scans with maxima at $\phi = \phi_B \pm \Delta \phi$, where $\phi_B$ is the Bragg angle expected for elastic scattering from the distorted vortex lattice  \cite{Rastovski2013,KuhnKFA,Kealey_2001}, and $\Delta \phi$ is the magnitude of the spin splitting of $\phi$.  Examples of the data are given in Fig.~\ref{Graph4}, together with schematic illustrations in Fig.~\ref{Graph4}e of the effect of the energy change on the scattering process.

This spin-splitting of the peaks, 2$\Delta\phi$ becomes more pronounced as $\Omega$ increases, because it is a function of $\Gamma_{\mathrm{VL}}$.  This arises because $2\Delta \phi=\left(2 k_0 / q_{\mathrm{VL}}\right)\left(\Delta \varepsilon / \varepsilon_0\right)$ where $q_{\mathrm{VL}}$ is the magnitude of the scattering vector along the minor axis i.e.~$q_{\mathrm{VL}} = q_{\mathrm{y}}$ as depicted in Fig.~\ref{Graph4}b,   $\Delta \varepsilon=\gamma \mu_N B$ and $\varepsilon_0=\hbar^2 k_0^2 / 2 m_n$, the neutron gyromagnetic ratio $\gamma=1.913$, the nuclear magneton $\mu_N=e \hbar / 2 m_n$ and $m_n$ is the neutron mass. This gives $\Delta\phi = C \,\Gamma_{\mathrm{VL}} \phi_B$ where $C = \gamma \sqrt 3 / 4 \pi = 0.2635$. We therefore have four peaks, two associated with the spot seen in the upper half of the detector in Fig.~\ref{Graph4}, with centers $\phi_{1/2}$ and two associated with the lower spot, $\phi_{3/4}$.  Their centres are at angles:
  \begin{align}
  \phi_{1/2} &= \phi_B \pm C \ \Gamma_{\mathrm{VL}} \ \phi_B  \\
  \phi_{3/4} &= -\phi_B \mp C \ \Gamma_{\mathrm{VL}} \ \phi_B 
  \end{align}

\noindent $\Gamma_{\mathrm{VL}}$ can then be extracted from the rocking curves by a simultaneous fit of the four peaks.  The individual peaks are treated as Lorentzians with a common integrated intensity and width under each field and $\Omega$ condition.  The relations between the peak centres are fixed by the equations above, giving two outputs: $\Gamma_{\mathrm{VL}}$ and the zero error in the $\phi$ motor positioning (found to be 0.09$^{\circ}$).  

\begin{figure}[t!]
        \includegraphics[width=\columnwidth]{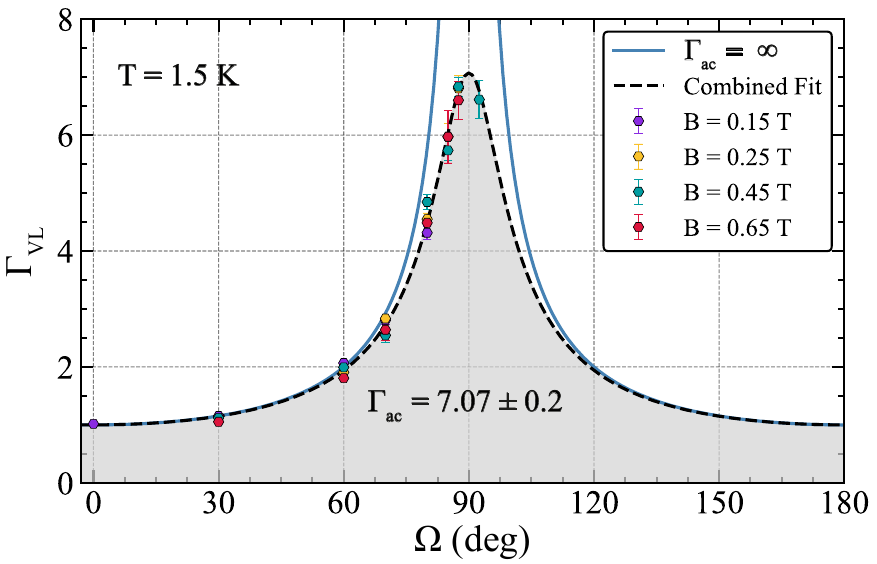}
	\caption{\textbf{Field-independent vortex lattice anisotropy}. The VL anisotropy measured at 1.5 K as a function of the applied field and the angle between the field and the \textbf{c} axis ($\Omega$). The dashed line shows the VL anisotropy calculated using Eqn.~\ref{gamma}, with $\Gamma_{ac}= 7.07$; the solid line is for $\Gamma_{ac}= \infty$.}
	\label{Graph5}
\end{figure}

All of these methods have been used to evaluate $\Gamma_{\mathrm{VL}}$ where possible, and they all agree within experimental error. The values obtained at each $\Omega$ are field independent. The weighted average of all methods at each condition is shown in Fig.~\ref{Graph5}. By fitting to Eqn.~\ref{gamma}, the superconducting anisotropy is found to be $\Gamma_{ac} = 7.07 \pm 0.2$.  

For anisotropic superconductors, like 2$H$-NbS$_2$, the variations in the penetration depth ultimately arise due to differences in the Fermi velocities and effective masses of the carriers in the different directions. $\Gamma_{ac}$ can therefore also be observed in the ratios of the superconducting coherence length, $\xi_{\mathrm{ab}} / \xi_{\mathrm{c}}$, and the upper critical fields, $B_{c 2}^{a b} / B_{c 2}^c$.  The upper critical fields have been measured by transport at 2 K, giving anisotropies ranging from 6.5 to 8.1 \cite{Onabe1978,Huang2022}, and a temperature dependent study of the heat capacity by Ka\ifmmode \check{c}\else \v{c}\fi{}mar\ifmmode \check{c}\else \v{c}\fi{}\'{\i}k \emph{et al.} \cite{Kacmarcik2010} finds a value of 7 for temperatures $0.3~T_c < T < T_c$.  While this value is in agreement with ours, Leroux \emph{et al.}~estimated a value of 11 from extrapolations of the penetration depth to 0 K \cite{Leroux2012}, while Cho \emph{et al.}~\cite{Cho_2022} used torque magnetometry to trace out the upper critical fields, combined with extremely precise in-plane angular alignment.  From this latter work, the in-plane upper critical field has an unusual temperature dependence.  Using the critical fields measured at 1.5 K gives $B_{c 2}^{a b} / B_{c 2}^c$ = 7.33.  However, this upper critical field appears to be significantly lower than the orbital upper critical field estimated from measurements close to $T_c$, which would give $\Gamma_{ac}$ $\sim$ 15.  Bi \emph{et al.}~\cite{Bi-2Dsc-orbital} have pointed out that surface superconductivity (up to $B_{c3} = 1.695 B_{c2}$) may be playing a role here.

In multi-band superconductors, the different bands may have different anisotropies, leading to field- and temperature-dependent superconducting anisotropies; a classic example is MgB$_2$ \cite{Cubitt-MgB2}.  2$H$-NbS$_2$ is widely accepted to be a two-band superconductor, but we do not observe any field dependence in our superconducting anisotropy, as was the case in the heat capacity studies of Ka\ifmmode \check{c}\else \v{c}\fi{}mar\ifmmode \check{c}\else \v{c}\fi{}\'{\i}k \emph{et al.} \cite{Kacmarcik2010}.  To engage with this question further, we first need to extract more information about the characteristic superconducting lengths from our data.

\subsection*{Integrated intensity and the field-dependent form factor}

For a structurally two-dimensional, well-ordered VL, the spatial variation of the magnetic field in the mixed state can be written as a Fourier series with components at different momentum transfers.  The Fourier component associated with the first order Bragg reflections from the diffracted lattice is referred to as the `form factor', and can be related to the integrated intensity $I(\textbf{q}_i)$ of a Bragg reflection from VL domain $i$ via the Christen formula \cite{Christen1977}: 

\noindent
\begin{equation}
I(\textbf{q}_i) = 2\pi V_i S \big(\frac{\gamma}{4} \big)^{2} \frac{\lambda_{n}^{2}}{\Phi_{0}^{2}q_i\cos(\zeta)} |F(\textbf{q}_i)|^{2}.
\label{FF}
\end{equation}
\noindent 

\noindent Here, $ V_i $ is the volume of the sample occupied by the VL domain $i$ (with $ V_{\textrm{I}} + V_{\textrm{II}} = V$, the total sample volume). $S$ is the incident neutron flux density (extracted through a measurement of the direct beam with known aperture size), \( \lambda_n \) is the neutron wavelength, \( \gamma \) is the gyromagnetic ratio of the neutron, \( \Phi_0 \) is the flux quantum, $\textbf{q}_i$ is the magnitude of the scattering vector for the relevant Bragg spot in the diffraction pattern, and \(\cos(\zeta\)) is the Lorentz factor, where \(\zeta \) is the angle between the reciprocal lattice vector  $\textbf{q}_i$ and the direction that is at right angles to the rocking axis \cite{squires_2012}.  For each domain, the relevant integrated intensities of individual Bragg peaks (see Methods) are averaged to give $I(\textbf{q}_i)$ for that domain; this is related to the form factor via Eqn.~\ref{FF}. 

\begin{figure*}[ht]
	\includegraphics[width=1\linewidth]{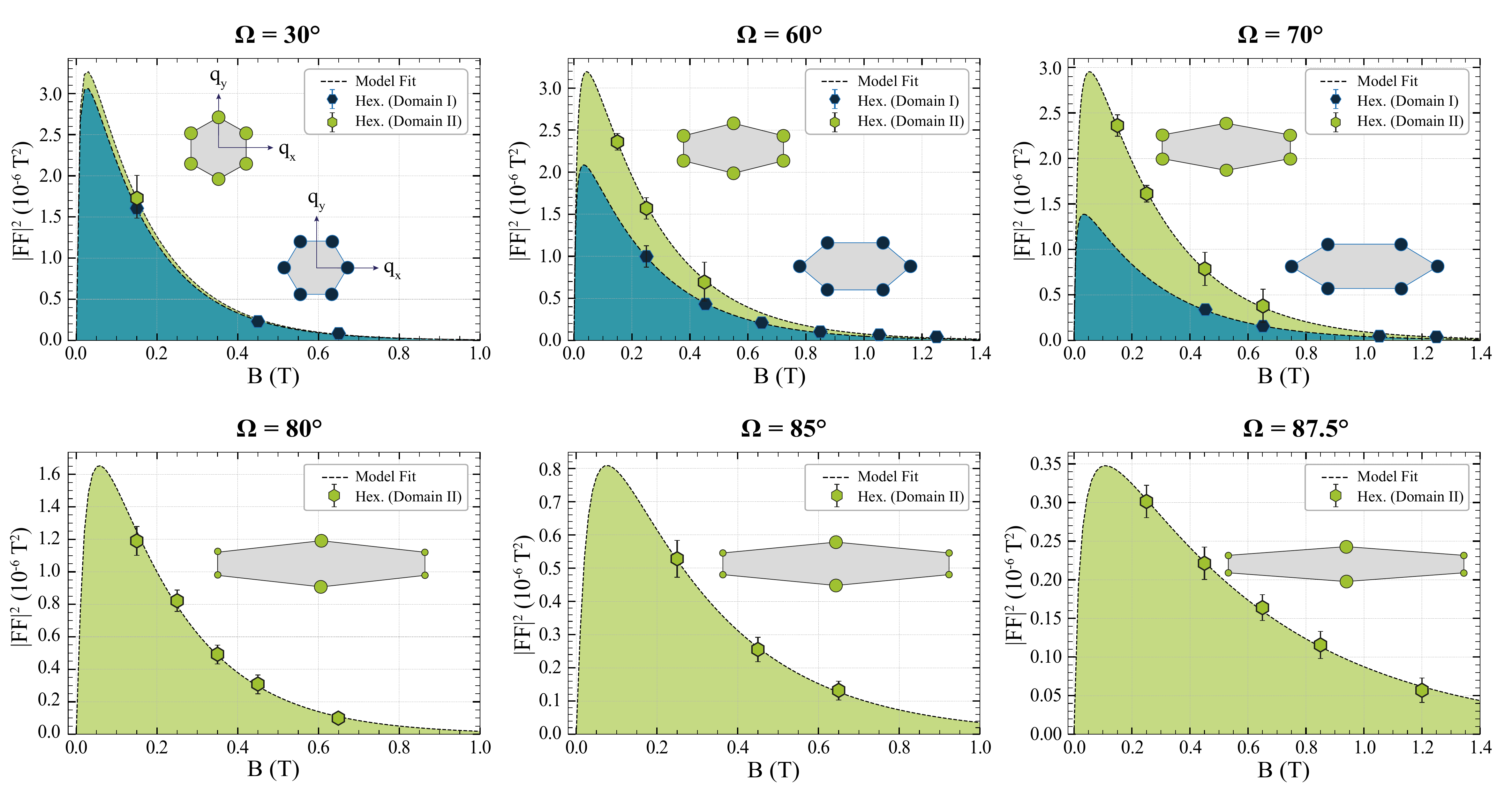}
	\caption{\textbf{Field dependence of the vortex lattice form factor for different field-sample angle $\Omega$}.
 Each panel shows the vortex lattice form factor as a function of the applied magnetic field $B$, measured at 1.5 K for a particular value of $\Omega$, ranging from $30^\circ$ to $87.5^\circ$. For the upper panels, scattering from two distinct hexagonal vortex lattice domains was present, denoted as Domain I (\protect\ffullhexagonSymbolOne) and Domain II (\protect\ffullhexagonSymbolTwo), while for the lower panels, the scattering arises only from Domain II. The data points represent experimental measurements of $|F(\textbf{q}_i)|^{2}$ with their respective error bars (see Supplementary Note 1).  The dashed black lines are the results from fits to Eqn.~\ref{eq:fq}, using theoretical values for the scattering vectors \textbf{q} of the Bragg peaks assuming $\Gamma_{ac} = 7$. Hexagons are depicted for each $\Omega$ to visualize the vortex lattice, constructed based on the actual anisotropy ratio $\Gamma_{\text{VL}}$($\Omega$).}
		\label{Graph6}
\end{figure*}

In an isotropic superconductor, the flux lines lie parallel to the applied field, with the screening supercurrents perfectly perpendicular to the direction of the average field, so that the local fields are all parallel to the applied field.  The form factor of the vortex lattice therefore contains only components parallel to the field ($b_z$ in Fig.~\ref{Graph1}c).  In 2$H$-NbS$_2$, this case applies if the field is perfectly parallel to the \textbf{c} axis. As soon as the field rotates away from this, transverse field components ($b_{x,y}$) will develop, leading to the spin-split scattering discussed above. This happens because the supercurrents tend to flow within the `easy' basal plane.  Whereas the flux lines still follow the average field direction, the form factor arises from spatially varying contributions from both the longitudinal and transverse field components. This has been fully described using anisotropic London theory by Kogan \cite{Kogan_1981} for uniaxial crystal systems like 2$H$-NbS$_2$, and expanded to account for the effective mass anisotropy of the carriers by Thiemann \emph{et al.} \cite{S.Thiemann_1989}. Kealey \emph{et al.} applied this practically to the biaxial superconductor YBa$_2$Cu$_3$O$_{7-\delta}$ \cite{Kealey_2001}, correcting some misprints in Ref.~\cite{S.Thiemann_1989}, but were only able to find qualitative agreement, as was the case in studies on Sr$_2$RuO$_4$ \cite{Rastovski2013,KuhnSRO_2017} and KFe$_2$As$_2$ \cite{KuhnKFA}. 
This may be due to difficulties in handling field-dependent superconducting anisotropies.  In 2$H$-NbS$_2$, we have a fixed value of $\Gamma_{ac}$, and we are able to validate the model of Thiemann \emph{et al.} over almost the entire angular range at multiple fields.

In the mixed state, the magnetic field distribution $\mathbf{B}(\mathbf{r})$ can be decomposed into a Fourier series over the set of reciprocal space wave vectors $\mathbf{q}$:
\begin{equation}
\mathbf{B}(\mathbf{r})=\sum_{\mathbf{q}} \mathbf{b}(\mathbf{q}) \exp (i\mathbf{q} \cdot \mathbf{r})
\end{equation}
\noindent The average field is in the $z$-direction, parallel to the applied field, and the  transverse fields $b_x(r)$ and
$b_y(r)$ are in the $xy$ plane.  For a given Bragg reflection at $\bf{q}$ = $(q_x,q_y)$, the field components are \cite{S.Thiemann_1989}:

\begin{align}
b_x & = \frac{(\lambda^2 m_{x z} q_{y}^2) B}{d} \label{eq:bx} \\
b_y & = \frac{(-\lambda^2 m_{x z} q_{x} q_{y}) B}{d} \label{eq:by} \\
b_z & = \frac{(1+\lambda^2 m_{z z} q^2) B}{d} \label{eq:bz}
\end{align}
\noindent where 
\begin{align}
d= & \left(1+\lambda^2 m_{y y} q_x^2+\lambda^2 m_{x x} q_y^2\right)\left(1+\lambda^2 m_{z z} q^2\right) - \lambda^4 m_{x z}^2 q^2 q_y^2.
\end{align}

\noindent Here $B$ is the average field and $\lambda=\left(\lambda_{a b}^2 \lambda_c\right)^{1 / 3}$ is the geometric mean of the penetration depths in the $a b$ plane and along the $\textbf{c}$ axis, and $m_{ij}$ are the tensor components of the effective mass of the charge carriers, referred to the $x$, $y$ and $z$ axes defined in Fig.~\ref{Graph1}c. The normalized effective masses along the unit cell axes are $m_a$, $m_b$, and $m_c$. For 2$H$-NbS$_2$, $m_a=m_b<m_c$, so $m_a{ }^2 m_c=1$. The superconducting anisotropy can be quantified as the ratio of the normalized effective masses, $\Gamma_{ac}=(m_c / m_a)^{1/2}$, so we can rewrite the effective mass components in the vortex frame into functions of $\Gamma_{ac}$ and $\Omega$:

\begin{align}
m_{x x} & =\Gamma_{a c}^{-2 / 3} \cos ^2 \Omega+\Gamma_{a c}^{4 / 3} \sin ^2 \Omega, \\
m_{y y} & =\Gamma_{a c}^{-2 / 3}, \\
m_{z z} & =\Gamma_{a c}^{-2 / 3} \sin ^2 \Omega+\Gamma_{a c}^{4 / 3} \cos ^2 \Omega, \\
m_{x z} & =\left(\Gamma_{a c}^{-2 / 3}-\Gamma_{a c}^{4 / 3}\right) \sin \Omega ~ \cos \Omega.
\end{align}
\noindent With these expressions we can then calculate the theoretical form factor, $F(\mathbf{q})$, for a given Bragg reflection, including a Gaussian cutoff term to account for the finite size of the vortex core \cite{Yaouanc_1997}: 
\begin{align}
F(\mathbf{q}) = (b_x^2 + b_y^2 + b_z^2)^{1/2}  \exp[-c (q_x^2 \xi_{\perp}^2) + q_y^2 \xi_{ab}^2)].
\label{eq:fq}
\end{align}

\noindent $\xi_{\perp}= [\xi_{ab}^2 \cos^2 \Omega + \xi_c^2 \sin^2 \Omega]$ represents the core width along the $q_x$ direction. Here,  $\xi_{\perp}$ is $\xi_{c}$ when $\Omega$ =  90$^{\circ}$ i.e.~{\bf B} $\parallel$ crystal planes. $\xi_{ab}$ represents the in-plane coherence length while $\xi_{c}$ is the coherence length for the \textbf{c}-axis direction. The constant $c$ is a core cutoff parameter; a quantitative comparison of this model and the numerical solution of the Eilenberger equations indicates that the most suitable value is $c$ = 0.44 \cite{Campillo2022_YBCO}.  


\begin{table*}
\begin{center}
   \renewcommand{\arraystretch}{1.5} 
   \begin{tabular}{M{0.15\textwidth} | M{0.15\textwidth} | M{0.15\textwidth} | M{0.15\textwidth} | M{0.15\textwidth}}
      \hline\hline
      \cellcolor{gray} \textcolor{white}{\textbf{$\boldsymbol{\Omega}$ ($^{\circ}$)}} & \cellcolor{gray} \textcolor{white}{\text{$\lambda_{\textbf{GM}}$ (nm)}} & 
      \cellcolor{gray} \textcolor{white}{\text{$\lambda_{\textbf{ab}}$ (nm)}} &  \cellcolor{gray} \textcolor{white}{\text{$\lambda_{\textbf{c}}$ (nm)}} &
      \cellcolor{gray} \textcolor{white}{\text{$\xi_{\textbf{ab,eff}}$ (nm)}} \\
      \hline
         \textbf{30}     & 277   $\pm$   5    &  145   $\pm$  3    &  1022   $\pm$  28  & 19.5   $\pm$  0.5     \\ \hline 
         \textbf{60}     & 275   $\pm$   3    &  144   $\pm$  2    &  1014   $\pm$  22  & 20.1   $\pm$  0.2     \\ \hline 
         \textbf{70}     & 270   $\pm$   2    &  141   $\pm$  1    &  995    $\pm$  21  & 23.2   $\pm$  0.4     \\ \hline 
         \textbf{80}     & 273   $\pm$   7    &  142   $\pm$  4    &  1006   $\pm$  32  & 34     $\pm$  1       \\ \hline 
         \textbf{85}     & 272   $\pm$   10   &  142   $\pm$  5    &  1001   $\pm$  40  & 34     $\pm$  1.5     \\ \hline 
         \textbf{87.5}   & 273   $\pm$   3    &  142   $\pm$  1    &  1006   $\pm$  21  & 24.6   $\pm$  0.4     \\ \hline\hline
   \end{tabular}
   \label{fit_results}
   \caption{\textbf{Fitted penetration depth and coherence length values at each angle $\Omega$}. These were obtained by fitting the field dependence of the form factor data measured at 1.5 K with the anisotropic London model with core-size correction factor described in Eqn.~\ref{eq:fq}.}
\end{center}
\end{table*}

The theoretical form factors depend on sample properties $\lambda$, $\xi_{ab}$ and $\xi_c$.  We fitted the field-dependent data separately for different values of $\Omega$.  The fits are very insensitive to the value of $\xi_c$, as it only influences the Domain I spots, and even then its contribution varies as $\xi^2_c\sin^2\Omega << \xi^2_{ab}\cos^2\Omega$, and so has little effect.  Accordingly, we fixed this parameter as $\xi_c = \xi_{ab} / \Gamma_{ac}$, and let $\lambda$ and $\xi_{ab}$  vary as a fittable parameters. Where two VL domains were present, their data were fitted simultaneously, with one additional (field-dependent) fitting parameter representing the fractions of the sample volume occupied by the two domains: $V_{\textrm{I}}/V$ and $V_{\textrm{II}}/V$. In Fig. ~\ref{Graph6}, we show the VL form factor variation as a function of field at $T$ = 1.5 K for six values of $\Omega$. At each field, we distinguish between the form factor values obtained from different types of Bragg spots, as these different spots will have different amounts of longitudinal and transverse field, so that the Domain I spots typically have lower form factors than  Domain II (illustrated more completely in Appendix A).  Each value represents the average taken over equivalent spots for a given domain.  The $(V_i/V) |F(\mathbf{q}_i) |^2$ were calculated from the experimental data using Eqn.~\ref{FF} and then fitted to Eqn.~\ref{eq:fq} using two different methods.

In the first approach, the intensities were fitted using the theoretical values for the scattering vector $\mathbf{q}$ associated with the peak given that $\Gamma_{ac} = 7$.  For Domain I ( \hexagonSymbolOne ), $q_x = (q_{\vartriangle}/2) ({ \Gamma_{\mathrm{VL}})^{1/2}  } $ and $q_y= (q_{\vartriangle}/2)  ( { 3 / \Gamma_{\mathrm{VL}}  })^{1/2}   $, while for Domain II ( \hexagonSymbolTwo ) $q_x=0$ and $q_y= (q_{\vartriangle})  / ({\Gamma_{\mathrm{VL}} })^{1/2} $.  In the second approach, we used the experimentally measured values for $q_x$ and $q_y$.  These two approaches yielded statistically indistinguishable results, given the experimental errors (an example is shown in Appendix B). This consistency increases confidence in the robustness of our findings.  In Fig.~\ref{Graph6} only the first approach is presented, as smooth fit lines can be calculated from Eqn.~\ref{eq:fq} using the theoretical Bragg spot positions.

The fits show excellent agreement with the description of the anisotropic superconductor given by Thiemann \emph{et al.} \cite{S.Thiemann_1989}, indicating that the tendency for supercurrents to flow in the $ab$-plane can be described in this way.  The values obtained for $\lambda$ and $\xi$ are given in Table II. 

Considering $\xi$ first, ideally, the in-plane coherence length $\xi_{ab}$ should represent the core size of the vortices, in which case it can be related to the upper critical field if the superconductivity is destroyed by orbital overlap, using the Ginzburg-Landau expression $B_{c2}^{\parallel c}$(0 K) = $\Phi_0 / 2\pi \xi_{ab}^2$.  By extrapolating to 0 K, its value can be estimated from the zero-temperature upper critical field $B_{c2}^{\parallel c}$.  This upper critical field is well documented for NbS$_2$, and for our sample, it is 1.8 T \cite{Majumdar_2020}.  If this is the orbital limit, then $\xi_{ab}$ (0 K) $\sim$ 13.5 nm.  Our fit results vary between 19.5 and 34 nm in size for this parameter, and so we consider that our measured value is an effective value, rather than the intrinsic value.  This could be because of flux pinning affecting the regularity of the vortex lattice.  This can be represented by a `static Debye-Waller factor' which arises from static disorder within the vortex lattice which can include local wiggling of the vortices \cite{Tinkham_1996}, or zigzagging of the vortices between basal planes at higher values of $\Omega$.  While we could not resolve $\xi_c$ in our fits, we note that using $B_{c2}^{\perp c}$(0 K) = $\Phi_0 / 2\pi \xi_{ab} \xi_{c}$, if we take the orbital value for the upper critical field of 24 T, as calculated by Cho \emph{et al.} \cite{Cho_2022} based on the Werthamer-Helfand-Hohenberg model, this gives $\xi_c \sim$ 1 nm. 

We now turn to the penetration depth, where the geometric mean at all values of $\Omega$ is found to be angle independent, as expected, with an average value of $272.3 \pm 1.3$ nm.  From this, $\lambda_{ab}$ = 141.9 $\pm$ 1.5 nm and $\lambda_c$ = 1003 $\pm$ 20 nm.  From the literature, there is sparse agreement: $\lambda_{ab}$ values between 83 nm \cite{Leroux2012} and 590 nm \cite{Tulapurkar-peakeffect} are reported. On our samples, a value for $\lambda_{ab}$ of 131 nm was obtained by Majumdar \emph{et al.}~ \cite{Majumdar_2020}.

Recently, Kogan \emph{et al.}~\cite{Kogan_lambda_2020}  established a relationship between the zero-temperature penetration depth  $\lambda (0)$ and the slope of the penetration depth $\lambda^{-2}(T)$ near $T_{\rm c}$, using a calculation analogous to the Helfand-Werthamer relationship between the zero-temperature upper critical magnetic field and its slope at $T_{\rm c}$. With further thermodynamic information, they obtained:

\begin{equation}
\lambda^{2}(0) \approx \left| \left(\frac{dH_{c2}}{dT}\right)_{T_c} \right| \frac{1}{T_c \gamma},
 \end{equation}

\noindent where $({dH_{c2}}/{dT})_{T_c}$  is the slope of the upper critical magnetic field $H_{c2}$ with respect to temperature ($T$) at $T_{\rm c}$, $T_{\rm c}$ is the critical temperature of the superconductor in question and $\gamma$ is the specific heat coefficient per unit volume.  Kogan \emph{et al.}'s model has been developed for isotropic $s$-wave superconductivity with non-magnetic scattering, and was successfully cross-checked against experimental values for $\lambda$ in V$_3$Si and Nb$_3$Sn. 

Using this model we have calculated $\lambda_{ab}$(0) $\approx 142.9$ nm for 2$H$-NbS$_2$, taking as inputs  $T_{\rm c} = 5.5~K$, $(dH^c_{c2} / dT)_{T=T_c} \approx - 0.25 \times 10^4 \ \text{Oe/K}$ ~\cite{Kacmarcik2010, Huang2022, Cho_2022}, and ~ $\gamma = 0.494 \times 10^4 \ \text{erg/cm}^3 \text{K}^2$~\cite{Hamaue_1986}.  This is in excellent agreement with our $\lambda_{ab}$ (1.5 K) as extracted from the fit. 

On the face of it, this is surprising, as 2$H$-NbS$_2$ is an anisotropic two-band superconductor, which also makes the field-independent superconducting anisotropy unexpected.  However, as Kogan \emph{et al.} noted, their results should still be applicable if the order parameter is constant over a Fermi surface of any given shape.  Indeed, Kogan \emph{et al.} tested their approach successfully on the two-band superconductor MgB$_2$, calculating $\lambda$(0) $\approx$ 176 nm, as compared to reported values of 180 - 185 nm.  

We therefore suggest that over the field range explored here at 1.5 K, there are no major changes in the contributions from the two gaps, with the dominant response coming from the large-gap band in 2$H$-NbS$_2$, which has predominantly 2D character, as proposed by Ka\ifmmode \check{c}\else \v{c}\fi{}mar\ifmmode \check{c}\else \v{c}\fi{}\'{\i}k \emph{et al.}~\cite{Kacmarcik2010}.  Interestingly, Noat \emph{et al.}~\cite{Noat_2015} have suggested that, in reality, only the large gap sheets (the cylinders centred on the K points) are intrinsically superconducting, and that the other gap develops parasitically by coupling through the pancake-like sheet coming from the chalcogen $p$-bands.  

\section*{Conclusions}

We have measured neutron diffraction by the vortex lattice in 2$H$-NbS$_2$ as a function of field angle and field magnitude at 1.5 K, and find that in fields up to 1.25 T, the intensity of the observed diffraction from the vortex lattice can be described extremely well by the Thiemann model using anisotropic London theory \cite{S.Thiemann_1989}, with the addition of a core-correction factor to account for the finite size of the vortex core.  This forms the first full validation of this model.  This process is aided by our clear observation that the superconducting anisotropy is constant and field-independent at $\Gamma_{ac} \sim 7$.  From this, we extract values for the London penetration depth of $\lambda_{ab} = 141.9 \pm 1.5 $ nm and $\lambda_{c} \sim$ 1 \textmu m.  For $\lambda_{ab}$, this experimental result fits with the recent model developed by Kogan \emph{et al.}~\cite{Kogan_lambda_2020}.

The dominant Fermi sheets in these conditions are expected to be the cylinders around the K point in reciprocal space, which hold most of the density of states at the Fermi surface, and are highly two-dimensional.  Coupling between these sheets and the more three-dimensional cylinders around the $\Gamma$ point therefore controls the overall superconducting anisotropy.  The interband coupling for 2$H$-NbS$_2$ has been found by Noat \emph{et al.}~\cite{Noat_2015} to be half that in 2$H$-NbSe$_2$, from fitting tunnelling spectroscopy data.  It is therefore not surprising that 2$H$-NbS$_2$ shows a stronger 2D character with a superconducting anisotropy of $7.07 \pm 0.2$, as compared to the value of $3.2 \pm 0.2$ measured by Gammel \emph{et al.}~\cite{Gammel1994} for 2$H$-NbSe$_2$ using the same method presented here. 

\section*{Methods}

Thin platelets of 2$H$-NbS$_{2}$ samples were grown using the chemical vapour transport technique, yielding high-quality single-crystals with optically flat surfaces on the macroscopic scale. A detailed description of this process for growing TMD crystals is given by Chareev \emph{et al.}~\cite{Chareev_2020}. The sample quality was confirmed by specific heat measurements at zero field, wherein a sharp jump centered at the superconducting transition temperature $T_{\rm c}$ = 5.5 K is seen.  This is considered to be the most reliable way to check sample quality \cite{Witteveen2021}, and matches well with the behaviour reported for other samples grown in the same laboratory \cite{Cho2021}.

The 2$H$-NbS$_{2}$ polytype is hexagonal, with space group $P6_3/mmc$ and lattice parameters $a = b \sim 3.31$ \AA~ and $c \sim 11.86$ \AA~\cite{Witteveen2021}. The platelets grow with the \textbf{c} axis normal to the platelet surface.  A mosaic of nine co-aligned crystals was made by mounting the platelets onto two aluminum sheets, so that all the crystal faces lay within an area  $10 \times 10 \, \text{mm}^2$. The mosaic had a total crystal thickness of 80 \textmu m with a total mass of $42.3 \, \text{mg}$. The in-plane alignment (along the $\textbf{a}^{*}$ - and $\textbf{b}$ - axes) was carried out using an optical microscope, with the clearly visible crystal hexagonal facets serving as points of reference. Overall, the misalignment of the $\textbf{a}^{*}$/$\textbf{b}$ - axes was measured to be less than 1$^{\circ}$.  

To diffract off the vortex lattice, small angle neutron scattering measurements were carried out at the D33 instrument \cite{Dewhurst:ks5488} at the Institut Laue-Langevin, Grenoble, France. The incident neutrons had wavelength $\lambda_{n}$ = 10 \AA, collimation of 10.3 m and a wavelength spread $\Delta\lambda_{n}/\lambda_{n}$ = 10\% full-width half-maximum (FWHM). The scattering patterns were collected on a (256$\times$128 pixels) two-dimensional position-sensitive multi-detector placed 10.0435 m after the sample, which was mounted in a 9 T horizontal-field cryomagnet.  The crystals were aligned with the $\textbf{a}^{*}$ axis vertical and the $\textbf{b}$ axis horizontal. A schematic of the arrangement of a typical SANS instrument when used for VL studies is shown in Fig.~\ref{Graph1}(a).

To create the vortex lattice, a magnetic field was applied. As shown in Fig.~\ref{Graph1}(b), the magnetic field was applied essentially parallel to the incoming neutron beam. The advantage of this parallel field geometry is that by rocking the sample and magnet together through small rocking angles about axes perpendicular to the neutron beam, denoted $\phi$ and $\Omega$, the diffraction conditions for all of the VL Bragg reflections can be met to give diffraction peaks on the 2D detector. In the work presented here, the sample and magnet were rocked about the horizontal axis perpendicular to the beam direction, i.e.~the $\phi$ angle.  At $\Omega = 0^{\circ}$, the sample and magnet were also rocked along the $\Omega$ angle.

To study the vortex lattice through the entire bulk of the sample, a relatively well-ordered vortex lattice needs to be prepared.  To do this, the sample was field cooled from above $T_{\rm c}$ to $T$ = 1.5 K while oscillating the magnitude of the applied magnetic field by ${\pm }$  1 \%  about the desired value. This procedure improves the orientational ordering of the VL, by keeping the vortices away from local pinning potentials \cite{Marziali2015}. This was done for magnetic fields over the range 0.15 T to 1.25 T.

To probe the superconducting anisotropy of 2$H$-NbS$_2$, we also rotated the magnetic field away from the $\textbf{c}$ axis towards the $\textbf{b}$ direction by an angle $\Omega$.  These rotations were all carried out above $T_{\rm c}$.  The experimental geometry is shown schematically in Fig.~\ref{Graph1}(c). As $\Omega$ is changed, the profile of the sample with respect to the beam changes, and so different sample apertures can be used at different angles; ideally only the sample volume should be illuminated by the neutron beam.  For $\Omega = 0^{\circ}$, 30$^{\circ}$, 80$^{\circ}$ and\ 85$^{\circ}$, a 12 mm diameter circular aperture was used. For $\Omega = 60^{\circ}$ and 70$^{\circ}$, a 7H $\times$ 10V \, mm$^2$ rectangular aperture was used.  For $\Omega =8 7.5^{\circ}$, the aperture size was 3H $\times$ 10V mm$^2$. 

Background data were collected in the normal state at 6.5~K, using the same rotation and tilt angles as those of the `foreground' measurements at 1.5 K. The background was then subtracted from the foreground, leaving only the vortex lattice signal. The analysis was done using the software program GRASP \cite{Dewhurst}.  Experimentally, the integrated intensity of a Bragg peak from the VL is determined by counting the number of neutrons detected in the region of the detector where the peak appears as a function of the rocking angle. For monitoring the form factor, only \(\phi \) rocking scans were used, so the Bragg condition for spots close to the horizontal axis is not fully met. This means that for Domain I, the integrated intensities of the four diagonal Bragg spots, represented by \hexagonSymbolOne, are used, whereas for Domain II, only the top and bottom Bragg spots, \hexagonSymbolTwo, are included. For the form factor calculations, the correct region of the detector to include for each peak was tested iteratively to maximise the signal-to-noise ratio after the background subtraction. The resulting rocking curves were fitted to a Lorentzian function with the background set to zero to give $I(\textbf{q}_i)$.  For $\Omega \leq 70^{\circ}$, each Bragg spot gave one maximum in the rocking curve.  At higher angles, the spin-splitting effect led to two maxima in each rocking curve. The integrated intensity in these cases consists of the sum of the intensities obtained from the two peaks, since each represents half of the incident neutron flux, corresponding to one direction of the neutron spin.

\section*{Acknowledgements}
AA and EB acknowledge support from the Crafoord Foundation (no.~20190930) and the Swedish Research Council under Project No.~201804704.  AA, EMF and EB thank the Institut Laue-Langevin for travel support to attend the experiment.  The authors gratefully acknowledge the Institut Laue-Langevin for the allocated beamtime.


\section*{Competing interests}
The authors have no conflicts to disclose.

\section*{Materials and correspondence}
The experimental data collected at the Institut Laue-Langevin is publicly available \cite{ILLdata}.  Processed experimental data are available upon request to the corresponding authors.

\section*{APPENDIX A: Longitudinal and Transverse magnetic field components of the flux line lattice in the anisotropic 2$H$-NbS$_2$}

\begin{figure}[t!]
  \includegraphics[width=0.85\columnwidth]{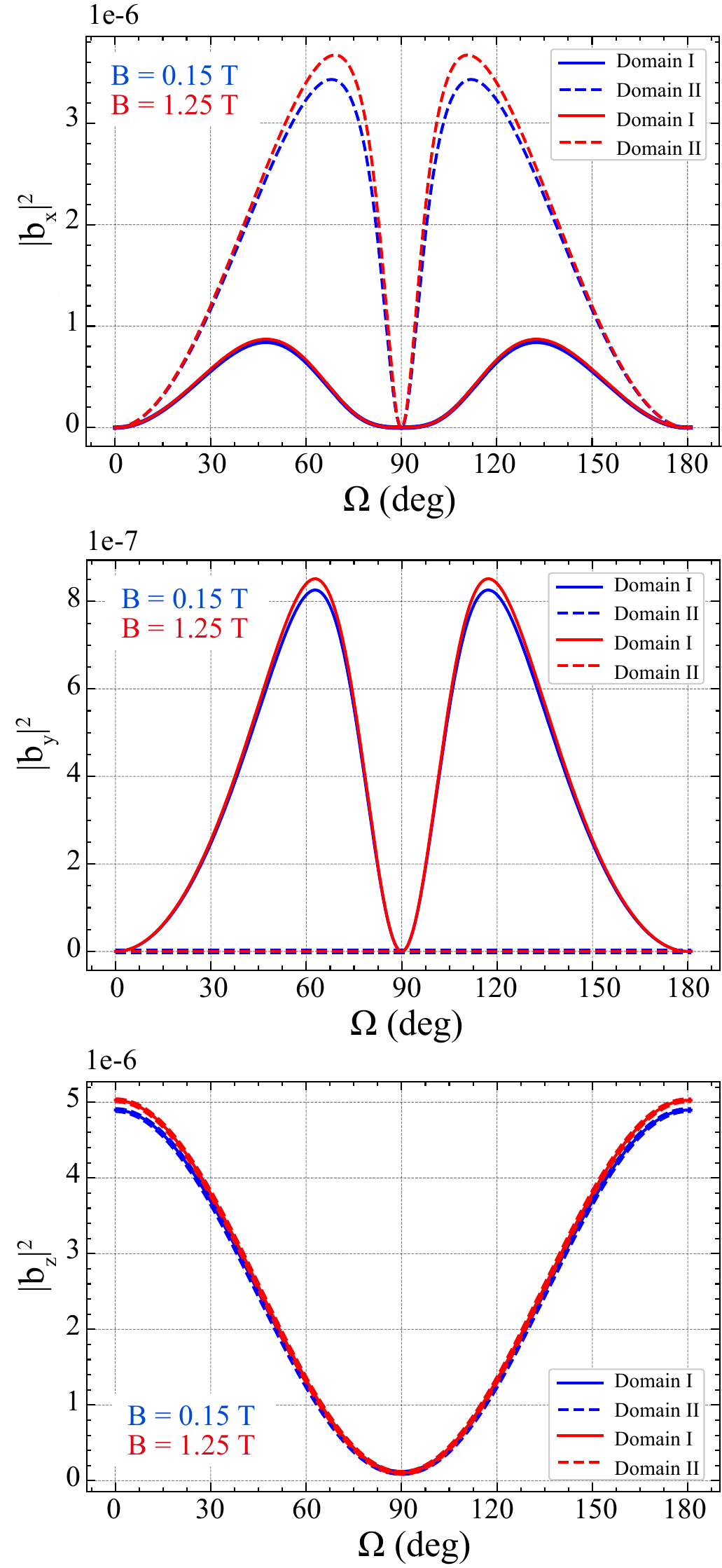}
 
\caption{Magnetic field components \(b_x\), \(b_y\), and \(b_z\) generating the form factor from the vortex lattice at the Bragg peaks \protect\hexagonSymbolOne from Domain I and \protect\hexagonSymbolTwo from Domain II, as a function of angle \(\Omega\) taken at 1.5 K for fields 0.15 T and 1.25 T.}
\label{Graph S1}
\end{figure}

As indicated in Fig.~\ref{Graph6},  
the Bragg reflections associated with Domain I have noticeably lower form factors than those for Domain II. This is due to the fact that there are two effects of the angular rotation on the VL form factor that bring this about. Firstly, the penetration depth is changing as $\Omega$ changes. Hence, the form factor of the observed Bragg spots on the left and right in the SANS patterns for both Domain I and Domain II (diagonal spots) decreases due to the longer penetration depth for currents in the {\bf c}-direction. However, this effect is insignificant for the top and bottom spots, whose form factor is primarily influenced by currents in the basal plane and remains relatively stable with small field rotations away from the c-axis. The second effect comes from $B_{\text{c2}}$, as the field is rotated away from the {\bf c} axis, the upper critical field $B_{\text{c2}}$ increases since it is larger when the field is closer to the basal plane. Consequently, a given magnetic field strength represents a smaller fraction of $B_{\text{c2}}$, reducing the core overlap effects in the superconductor. This reduction in core overlap is uniform for all diffraction spots in the SANS pattern. The combined result of these changes is an increase in the form factor for the top and bottom spots, making them more intense than the left and right spots, which diminish in intensity upon rotation away from the {\bf c}-axis. This specific outcome is notably observed at rotation angles of $80^{\circ}$, 85$^{\circ}$, and $87.5^{\circ}$ as shown in Fig.~2. 
In Fig.~\ref{Graph S1} we show how the form factor in the form of transverse and longitudinal field components $b_x$, $b_y$, and $b_z$ is changing as a function of both rotation angle ($\Omega$) and applied magnetic field ($B$). The calculations involve the diagonal spots of Domain I ( \hexagonSymbolOne) and only the top and bottom spots of Domain II ( \hexagonSymbolTwo) using $\Gamma_{ac}= 7$ and Eqns.~6-8. 

\section*{APPENDIX B: Field dependent form factor fit using the experimental values of $q_x$ and $q_y$ (Approach II)}

\begin{figure*}[ht]
\centering
\includegraphics[width=1\linewidth]{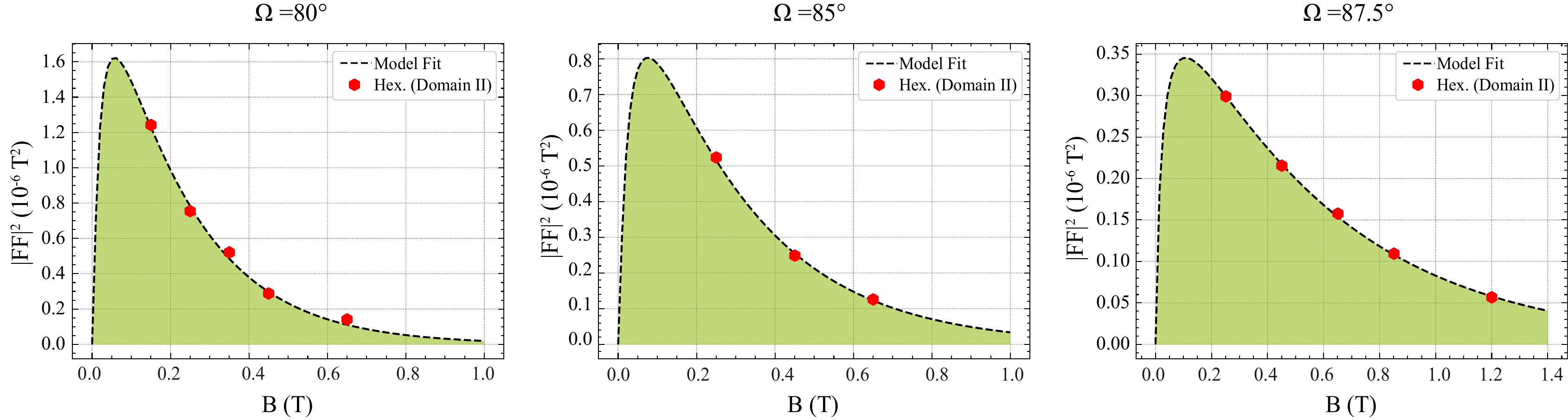}
\caption{ Vortex lattice form factor for angles $\Omega = 80^{\circ}$, 85$^{\circ}$ and $87.5^{\circ}$ as a function of the field at 1.5 K.}
\label{Graph S2}
\end{figure*}

In Fig.~\ref{Graph6}, we illustrated the fit of the form factor data using the theoretical definition for $q_x$ and $q_y$ (Approach I). Here, we show our Approach II by fitting the data with the experimentally measured values for $q_x$ and $q_y$. As shown in Fig~\ref{Graph S2}, we show the fit for angles $\Omega$ = $80^{\circ}$, 85$^{\circ}$ and $87.5^{\circ}$ as an example. The red data points are the results from the fitting process using Approach II for each angle $\Omega$. The black dashed lines are from the Approach I fit shown in Fig.~\ref{Graph6}. This shows that both fit procedures are robust and consistent, and the output fit from both approaches yields the same $\lambda_{GM}$,  $\xi_{ab}$, and $\xi_{c}$ within the experimental error.

%


\end{document}